\documentclass[10pt,english]{article}
\usepackage[latin9]{inputenc}
\usepackage[a4paper]{geometry}
\usepackage{verbatim}
\usepackage{amstext}
\usepackage{authblk}
\usepackage{graphicx}
\usepackage{cite}
\usepackage{hyperref}
\usepackage{sidecap}

\makeatletter

\begin{document}



\title{Asymptotic Theory of the Ponderomotive Dynamics of an Electron Driven
by a Relativistically Intense Focused Electromagnetic Envelope}

\author[1,2]{O.B. Shiryaev \thanks{shiryaev@kapella.gpi.ru,  \url{https://sites.google.com/site/drolegbshiryaev}}}

\affil[1]{Prokhorov General Physics Institute of the Russian Academy of Sciences,  
\break Vavilov Street 38, 117942, Moscow, Russia \url{http://www.gpi.ru/eng/index.php}}
\affil[2]{Medicobiologic Faculty, N.I. Pirogov Russian National Medical Research University, 
\break Ostrovitianov Street 1, Box 117997, Moscow, Russia \url{http://rsmu.ru/3784.html}}

 \maketitle

\begin{abstract}
A formalism for describing relativistic ponderomotive effects, which
occur in the dynamics of an electron driven by a focused relativisticaly
intense optical envelope, is established on the basis of a rigorous
asymptotic expansion of the Newton and Maxwell equations in a small
parameter proportional to the ratio of radiation wavelength to beam
waist. The pertinent ground-state and first order solutions are generated
as functions of the electron proper time with the help of the Krylov-Bogolyubov
technique, the equations for the phase-averaged components of
the ground state arising from the condition that the first-order solutions
sustain non-secular behaviour. In the case of the scattering of a
sparse electron ensemble by a relativistically intense laser pulse
with an axially symmetric transverse distribution of amplitude, the resulting
ponderomotive model further affords averaging over the random initial
directions of the electron momenta and predicts axially symmetric
electron scatter. Diagrams of the electron scatter directionality
relative to the optical field propagation axis and energy spectra
within selected angles are calculated from the compact ponderomotive
model. The hot part of the scatter obeys a clear energy-angle dependence
stemming from the adiabatic invariance inherent in the model, with smaller energies allocated
to greater angular deviations from the field propagation axis, while
the noise-level cold part of the scatter tends to spread almost uniformly
over a wide range of angles. The allowed energy diapasons within specific
angular ranges are only partially covered by the actual high-energy
electron scatter.
\end{abstract}


 \section{Introduction}
 
Interest in various aspects of the nonlinear dynamics of electrons
driven by electromagnetic fields deepened with the advent of relativistic
intensity laser physics, an area of study which took shape as attainable
laser intensities crossed the threshold of around 10$^{19}$
W$/$cm$^2$ \cite{MourouTajimaBulanov}. The quantity
is conventionally termed the relativistic intensity. At the extreme
intensities, laser pulses both induce relativistic motions of electrons
born in or injected into the focal spot and have the ability to relay
to them considerable post-interaction energies. The drift of an electron
across the focal spot of an intense electromagnetic envelope on a
timescale slow compared to the field cycle is essentially the
ponderomotive effect responsible, in particular, for the net energy
gain by the particle. Developing a formal description of the relativistic
ponderomotive dynamics of an electron in a superstrong focused electromagnetic
envelope is the purpose of the present study.

The baseline scenario behind the motions of an electron driven by
the Lorentz force exerted by a relativistically intense focused envelope
has been revealed in multiple simulations (e.g, see 
\cite{Hora1998,KongCapture,WangXoYuanvacuumAccelerat,Cao2002,PangCapture,Salamin2002,My-SingleElectron}).
An electron interacting with a relativistically intense laser pulse
gets captured by the field and draws energy from it to ultimately
get released with a certain residual energy.
The models employed in various pertinent studies comprise relativistic
Newton's equations coupled to the expressions describing the propagation
of focused optical envelopes in vacuum. The formulations for the field
range in complexity from those corresponding to a Gaussian transversely
polarized beam with a waist to more sophisticated models embracing
the corrections which account for the longitudinal component of the
field 
\cite{Lax,QuesnelMora} 
or, furthermore, a combination
of the latter with the pulse duration-related corrections to the transverse
component of the field 
\cite{MilantyevHighOrderCorrections,LPB-2017}.

The concept of ponderomotive dynamics of an electron in a high-frequency
envelope can be traced back to the elegant quasilinear analysis by
Gaponov and Miller \cite{GaponovMiller} which reveals a ponderomotive
force proportional to the gradient of the field intensity as the time-averaged
driver of the electron dynamics (also, see \cite{Kibble}). Multiple
attempts were made to generalize the ponderomotive force concept
to cover the relativistic intensity range as the attainable laser
intensities rose. The majority of studies replicated the approach of
\cite{GaponovMiller}, transforming the basic equations, adopting
a priori assumptions about the character of the oscillations of their
various terms including the relativistic mass factor of the electron,
and performing averaging in time on such basis 
\cite{Taranukhin,QuesnelMora,SerovPonderomotive}.
An effort to implement fully the Krylov-Bogolyubov technique can be
found in \cite{MilantyevQuasiRel}, but, in the practically important
case of the linear polarization of laser radiation, the study is limited
to the quasi-relativistic case due to the complexity of assessing
the mode of oscillations of the electron mass factor. 

An important study highlighted the crucial difference between the
averaging in time and averaging over the optical field oscillation phase in exploring
the dynamics of a laser-driven electron \cite{Tokman}. The conclusion
therein is that the latter approach, representing a departure from
the structural logic defined by \cite{GaponovMiller} and implemented
in the other studies cited above, becomes necessary if the longitudinal
displacements of the field-driven particle are appreciable, which
clearly is the case if the electron motion is powered by a focused
intense laser pulse. The averaging attempt in terms of proper time,
which is directly related to phase, was performed in \cite{NarozhniyFofanov},
but the study treated the impractical and relatively simple case of
circular polarization of the electromagnetic field and involved no
specific field description.

Several studies detailing the ponderomotive dynamics in the one-dimensional
case, corresponding to infinite focal spot size, are also
available 
\cite{DodinFischFraiman,DodinFisch}. 
It should be
borne in mind that in such geometry the electron dynamics is governed
by a simple invariant linking the transverse and longitudinal components
of the particle momenta. An analog of such invariance being reasonably
expected to manifest itself in the 3D situation \cite{Hartemann},
no theory for it is found in the above ponderomotive studies.

A formal asymptotic solution to the relativistic Newton's equation
for an electron driven by the Lorentz force, which is generated by
a linearly polarized intense focused optical envelope, is developed
in the present paper. The small parameter for asymptotic expansion
is $\epsilon=\lambda/\left(2 \pi w_{0}\right)$, where $\lambda$ and
$w_{0}$ are the laser pulse wavelength and waist. The role of an
independent variable within the implementation of the asymptotic algorithm
is assigned to the optical field phase, with the related electron
fast and slow proper times being introduced. The optical field expression
factoring into the Lorentz force includes high-order corrections in
$\epsilon$, which are indispensable to the resolution of the ground-state
asymptotic terms in the framework of the Krylov-Bogolyubov method.
As is customary in the framework of the latter, the nonosillatory
parts of the electron coordinates and momentum emerge as integration
constants in solutions to lower-order equations to be determined from
the conditions that secular terms must be absent in the solutions
to higher-order ones. The equations embodying this requirement ultimately
provide a closed problem for the nonoscillatory variables. It can
be interpreted as the corresponding set of averaged equations for
the electron dynamics, though the derivation employs no explicit averaging
or any a prioi assumptions concerning the character of oscillations
of the electron relativistic mass factor or other quantities. Importantly,
the lowest-order, fully three-dimensional approximation for the laser-driven
relativistic Newton's equation for an electron is shown to sustain
an adiabatic invariant which appears to be analogous in shape to the
well-known one found in one-dimensional geometry and, in the three-dimensional
case, links the energy and the angle of motion relative to the field
propagation axis for an electron ejected from the laser focal spot. 

The equations resulting from the asymptotic expansion acquire a relatively compact form
if the amplitude distribution of the optical field is axially symmetric.
In the reduced form, the equations warrant the conclusion that, for
a linearly polarized laser pulse, the scatter of an ensemble of electrons
with initial momenta having a uniform angular distribution in the
plane perpendicular to the laser propagation axis also spreads uniformly
over angles in the transverse plane. Solutions to the asymptotic equations
describing the scattering of a sparse ensemble of electrons by relativistically
intense laser radiation with axially symmetric amplitude distribution
are presented below and applied to delineate the characteristics of
the scatter.

\section{Asymptotic Solutions to Relativistic \\
Newton's Equations for a Laser-Driven Electron}

\noindent The nonlinear dynamics of an electron relativistically
driven by electromagnetic radiation obeys Newton's equations 
\begin{eqnarray}
\gamma\partial_{t}x&=&p_{x},\quad\gamma\partial_{t}y=p_{y},\quad\gamma\partial_{t}z=p_{z},  \label{eq:RelativisticMomenta} \\
\partial_{t}\mathbf{p}&=&\partial_{t}\mathbf{A}-\gamma^{-1}\left(\mathbf{p}\times(\nabla\mathbf{\times A})\right),\quad\gamma=\sqrt{1+\mathbf{p}^{2}}, \label{eq:NewtonEq}
\end{eqnarray}
\noindent where $\nabla=(\partial_{x},\partial_{y},\partial_{z})$,
$\gamma$ is the relativistic mass factor, and $\mathbf{A}$ stands
for the vector potential of the field propagating in vacuum. Assuming
that the field has the shape of a focused envelope, coordinates and
time are normalized by the focal spot size $w_{0}$ and by $w_{0}/c$
respectively, and the vector potential is normalized by $mc^{2}/e$.
The vector potential solves Maxwell's equations 

\[
\triangle\mathbf{A}-\partial_{t}^{2}\mathbf{A}=0, \quad
(\nabla,\mathbf{A})=0
\] 

\noindent
(Coulomb gauge) and, in the case of linear
polarization, has the asymptotic structure represented by 

\begin{eqnarray*}
A_{x}&=&\exp\left(i\theta\right)\left(a(\tau,x,y,s)+\sum_{m=1}^{\infty}\epsilon^{m}a_{xm}(\tau,x,y,s)\right)+c.c., \\
A_{y}&=&0, \quad
A_{z}=\exp\left(i\theta\right)\sum_{m=1}^{\infty}\epsilon^{m}a_{zm}(\tau,x,y,s)+c.c.,
\end{eqnarray*}

\noindent
where the variables defined as $\theta=(t-z)/\epsilon$ and $s=\epsilon\theta$
can be interpreted as fast and slow proper times of the electron.
Furthermore, $\tau=2\epsilon z$ and $\epsilon=\left(\lambda/2\pi w_{0}\right)$,
$\lambda$ denoting the radiation wavelength and $\epsilon$ being
a small parameter. As shown in \cite{LPB-2017}, the ground-state
and first-order results for the vector potential stem from the resulting asymptotic equations 

\[
-4i\partial_{\tau}a+\triangle_{\bot}a=0, \quad
-4i\partial_{\tau}a_{x1}+\triangle_{\bot}a_{x1}=4\partial_{\tau s}^{2}a,
\]

\noindent
and are of the form
\begin{eqnarray}
a(\tau,x,y,s)         & =& a_{0}(s)u(x,y,\tau),\label{eq:LowestOrderEnvelope} \\
a_{x1}(\tau,x,y,s) & =& ia_{0}^{\prime}(s)\partial_{\tau}\left(\tau u(x,y,\tau)\right),\label{eq:FirstOrderEnvelope} \\
a_{z1}(\tau,x,y,s) & =& -ia_{0}(s)\partial_{x}u(x,y,\tau),\label{eq:FirstOrderLongitudinalEnvelope}
\end{eqnarray}
\noindent where $u(x,y,\tau)$ is the field amplitude to be calculated from
the Schroedinger equation 

\[
-4i\partial_{\tau}u+\triangle_{\bot}u=0
\]

\noindent
with $\Delta_{\perp}=\partial_{x}^{2}+\partial_{y}^{2}$, and $a_{0}(s)$
describes the laser pulse temporal profile. The simplest corresponding
solution specifically treated in the concluding part of the present study
portrays a Gaussian pulse

\noindent 
\begin{eqnarray*}
u(x,y,\tau)&=&\frac{\Lambda(\tau,r)}{\sqrt{\tau^{2}+1}}\exp\left(i\psi(\tau,r)\right), \\
\Lambda(\tau,r)&=&\exp\left(-\frac{r^{2}}{\tau^{2}+1}\right),\quad \psi(\tau,r)=-\frac{\tau r^{2}}{\tau^{2}+1}+\arctan\tau
\end{eqnarray*}
with $r=\sqrt{x^{2}+y^{2}}$ (a generalization of the above solution
involving Laguerre modes in the ground state and the pertinent first-order
corrections is also available \cite{LPB-2017}). The objective at
hand is to develop asymptotic solutions in $\epsilon$ to Eqs. (\ref{eq:RelativisticMomenta})-(\ref{eq:FirstOrderLongitudinalEnvelope}),
that is, to solve the relativistic electron dynamics problem under
the same assumptions which yield the above envelope solutions to Maxwell
equations for the field in vacuum. It should be noted that, due to the architecture
of the ensuing asymptotic algorithm, the first-order corrections to
the field prove necessary for obtaining even the ground-state asymptotic
solutions. 


Consider the electron dynamics equations (\ref{eq:RelativisticMomenta})
and (\ref{eq:NewtonEq}) with the field vector potential originating
from Eqs. (\ref{eq:LowestOrderEnvelope})-(\ref{eq:FirstOrderLongitudinalEnvelope}).
Using the obvious fact that $\partial_{t}=(j/\epsilon\gamma)\partial_{\theta},$
where
\begin{equation}
j=\gamma-p_{z},\label{eq:AdiabaticInvariant}
\end{equation}
the problem is conveniently switched from $t$ to $\theta$ as the independent
variable. To develop an asymptotic solution in $\epsilon$, the variables
$s$ and $\theta$ may, in the process of cultivating the approximations,
be treated as independent. The asymptotic series for the electron
coordinates and momenta are
\begin{eqnarray*}
x(t)&=&x_{0}(s,\theta)+\epsilon x_{1}(s,\theta)+\ldots,\quad p_{x}(t)=p_{\text{x0}}(s,\theta)+\epsilon p_{\text{x1}}(s,\theta)+\ldots \\
y(t)&=&y_{0}(s,\theta)+\epsilon y_{1}(s,\theta)+\ldots,\quad p_{y}(t)=p_{\text{y0}}(s,\theta)+\epsilon p_{\text{y1}}(s,\theta)+\ldots \\
\tau(t)&=&\tau_{0}(s,\theta)+\epsilon\tau_{1}(s,\theta)+\ldots,\quad p_{z}(t)=p_{\text{z0}}(s,\theta)+\epsilon p_{\text{z1}}(s,\theta)+\ldots
\end{eqnarray*}

\noindent
In this framework, the ground-state results are obtained from ordinary
differential equations in $\theta$ and found to involve arbitrary
functions depending on $s$. These functions are to
be determined based on the requirement that the solutions to higher-order
equations remain free of secular growth. The ground-state solution
shown below is fully nonlinear and analogous to the well-known solution
to the one-dimensional problem. The higher-order equations are linear
ordinary differential equations in $\theta$, and their solutions are
explicitly spelled out in what follows with an eye to deriving non-secular
behaviour conditions without unwarranted averaging assumptions. The conditions for the absence of secular growth are identifiable
as the de facto averaged equations of the electron dynamics.


Denote $m(x,y,\tau)=\mathrm{Re}\,u(x,y,\tau)$ and $n(x,y,\tau)=\mathrm{Im}\,u(x,y,\tau)$.
The ground-state solutions for the coordinates and the transverse
momenta, as obtained by substituting the above series into Eqs. (\ref{eq:RelativisticMomenta})-(\ref{eq:NewtonEq})
and (\ref{eq:LowestOrderEnvelope})-(\ref{eq:FirstOrderLongitudinalEnvelope}),
are easily integrated with the result that some of the unknowns are
independent of the fast variable $\theta$, namely
\begin{eqnarray}
x_{0}(s,\theta)&=&x_{0a}(s),\quad y_{0}(s,\theta)=y_{0a}(s), \quad 
\tau_{0}(s,\theta)=\tau_{0a}(s),\label{eq:GroundStateCoords} \\
p_{\text{x0}}(s,\theta)&=&a_{0}(s)\left[m\left(x_{0a}(s),y_{0a}(s),\tau_{0a}(s)\right)\cos\theta
- 
n\left(x_{0a}(s),y_{0a}(s),\tau_{0a}(s)\right)\sin\theta\right]
+p_{\text{x0a}}(s),\\ \label{eq:GroundState-px}
p_{y0}(s,\theta)&=&p_{y0a}(s).\label{eq:GroundState-py} 
\end{eqnarray}
The functions $x_{0a}(s)$, $y_{0a}(s)$, $\tau_{0a}(s)$, $p_{\text{x0a}}(s)$,
$p_{y0a}(s)$, $j_{0}(s)$ emerge at this point as integration constants. In
a meaningful semblance to the 1D case, the above ground-state solutions
show that the quantity $j$ defined by Eq. (\ref{eq:AdiabaticInvariant}),
calculated to the lowest order, is independent of the fast proper
time $\theta$, namely,
\begin{equation}
\gamma_{0}(s,\theta)-p_{z0}(s,\theta)=j_{0}(s),\label{eq:AdiabaticInvariantGroundState}
\end{equation}
where $\gamma_{0}(s,\theta)=\sqrt{1+p_{x0}^{2}(s,\theta)+p_{y0}^{2}(s,\theta)+p_{z0}^{2}(s,\theta)}$,
while $j_{0}(s)$ also awaits being defined using higher-order approximations.
Therefore, we have

\begin{equation}
p_{\text{z0}}(s,\theta)=\frac{p_{\text{x0}}(s,\theta){}^{2}+p_{\text{y0}}(s,\theta){}^{2}-j_{0}(s){}^{2}+1}{2j_{0}(s)}.\label{eq:GroundStateLongitudinalMomentum}
\end{equation}
Accordingly, the electron mass factor in the ground state is 
\[
\gamma_{0}(s,\theta)=\frac{p_{\text{x0}}(s,\theta){}^{2}+p_{\text{y0}}(s,\theta){}^{2}+j_{0}(s){}^{2}+1}{2j_{0}(s)},
\]

\noindent so that $E=\gamma_{0}(s,\theta)-1$. 
In the first order, the solution to the last of Eqs. (\ref{eq:RelativisticMomenta}),
rewritten for the variable $\tau$, evaluates to 

\[
\tau_{1}(s,\theta)=\tau_{\text{1a}}(s)-\theta\tau_{0a}'(s).
\]

\noindent
The above first-order solution could exhibit secular behaviour in the sense that  it is a growing function of $\theta$ and can thus become comparable in magnitude to the ground state, which runs contrary to the assumptions underlying the asymptotic approach. Obviously, $\tau_{1}(s,\theta)$ stays secularity-free provided that $\tau_{0a}(s)$ in Eq. (\ref{eq:GroundStateCoords})
is a constant, so that eventually
\[
\tau_{0a}(s)=\tau_{0a},\quad\tau_{1}(s,\theta)=\tau_{\text{1a}}(s).
\]

\noindent Similarly, the higher-order solutions  obtained below generally involve terms which grow in $\theta$, and, in every case, the absence of secularity requirement dictates further conditions to be imposed on the parameters of the ground-states. 

The following notations are used below for brevity
\begin{eqnarray*}
m_{0}(s)=m(x_{0a}(s),y_{0a}(s),\tau_{0a}), \quad 
n_{0}(s)=n(x_{0a}(s),y_{0a}(s),\tau_{0a}), \\
m_{1}(s)=\frac{\partial m(x_{0a}(s),y_{0a}(s),\tau_{0a})}{\partial x}, \quad
 n_{1}(s)=\frac{\partial n(x_{0a}(s),y_{0a}(s),\tau_{0a})}{\partial x}, \\
m_{2}(s)=\frac{\partial m(x_{0a}(s),y_{0a}(s),\tau_{0a})}{\partial y},\quad
n_{2}(s)=\frac{\partial n(x_{0a}(s),y_{0a}(s),\tau_{0a})}{\partial y}, \\
m_{3}(s)=\frac{\partial m(x_{0a}(s),y_{0a}(s),\tau_{0a})}{\partial\tau}, \quad
n_{3}(s)=\frac{\partial n(x_{0a}(s),y_{0a}(s),\tau_{0a})}{\partial x}.
\end{eqnarray*}


\noindent The solutions to the first-order equations for the transverse coordinates
are 

\begin{eqnarray}
x_{1}(s,\theta)&=&\frac{a_{0}(s)\left(m_{0}\left(s\right)\sin\theta+n_{0}\left(s\right)\cos\theta\right)}{j_{0}(s)}
+ 
x_{1a}(s)
+ 
\left(\frac{p_{\text{x0a}}(s)}{j_{0}(s)}-x_{0a}'(s)\right)\theta,\label{eq:x1Solution}\\
y_{1}(s,\theta)&=&y_{1a}(s)+\left(\frac{p_{\text{y0a}}(s)}{j_{0}(s)}-y_{0a}'(s)\right)\theta.\label{eq:y1Solution}
\end{eqnarray}


The solutions to the first-order equations for the transverse components
of the electron momentum are

\begin{eqnarray}
p_{\text{x1}}(s,\theta)&=&\sum_{k=1,2}\left(\alpha_{x,k}(s)\cos(k\theta)
+  
\beta_{x,k}(s)\sin(k\theta)\right)
+ 
p_{x1a}(s)
-  \nonumber \\ &-&
\left(p_{\text{x0a}}'(s)+\frac{a_{0}^{2}(s)\left(m_{0}(s)m_{1}(s)+n_{0}(s)n_{1}(s)\right)}{2j_{0}(s)}\right)\theta,\label{eq:px1Solution} \\
p_{\text{y1}}(s,\theta)&=&\sum_{k=1,2}\left(\alpha_{y,k}(s)\cos(k\theta)+\beta_{y,k}(s)\sin(k\theta)\right)
+ 
p_{y1a}(s)
-  \nonumber \\ &-& 
\left(p_{\text{y0a}}'(s)+\frac{a_{0}^{2}(s)\left(m_{0}(s)m_{2}(s)+n_{0}(s)n_{2}(s)\right)}{2j_{0}(s)}\right)\theta,\label{eq:py1Solution}
\end{eqnarray}

\noindent where the coefficients of the oscillatory parts are 

\begin{eqnarray*}
\alpha_{x,1}(s)&=&a_{0}(s)\left(-\frac{n_{1}(s)p_{x0a}(s)}{j_{0}(s)}+m_{1}(s)x_{1a}(s)+m_{2}(s)y_{1a}(s)
+ 
\text{\ensuremath{m_{3}}(s)\ensuremath{\tau_{\text{1a}}}(s)}\right)-a_0'(s)\left(\text{\ensuremath{n_{0}}(s)+}n_{3}(s)\tau_{0a}\right),\\
\alpha_{x,2}(s)&=&a_{0}^{2}(s)\frac{m_{0}(s)n_{1}(s)+n_{0}(s)m_{1}(s)}{4j_{0}(s)}, \\
\beta_{x,1}(s)&=&-a_{0}(s)\left(\frac{m_{1}(s)p_{x0a}(s)}{j_{0}(s)}+n_{1}(s)x_{1a}(s)+n_{2}(s)y_{1a}(s)
+ 
\text{\ensuremath{n_{3}}(s)\ensuremath{\tau_{\text{1a}}}(s)}\right)-a_0'(s)\left(\text{\ensuremath{m_{0}}(s)+}m_{3}(s)\tau_{0a}\right), \\
\beta_{x,2}(s)&=&a_{0}^{2}(s)\frac{m_{0}(s)m_{1}(s)-n_{0}(s)n_{1}(s)}{4j_{0}(s)}, \\
\alpha_{y,1}(s)&=&-a_{0}(s)\frac{n_{2}(s)p_{x1a}(s)}{j_{0}(s)},\quad
\alpha_{y,2}(s)=-a_{0}^{2}(s)\frac{n_{0}(s)m_{2}(s)+m_{0}(s)n_{2}(s)}{4j_{0}(s)}, \\
\beta_{y,1}(s)&=&-a_{0}(s)\frac{m_{2}(s)p_{x1a}(s)}{j_{0}(s)}, \quad
\beta_{y,2}(s)=a_{0}^{2}(s)\frac{n_{0}(s)n_{2}(s)-m_{0}(s)m_{2}(s)}{4j_{0}(s)}. \\
\end{eqnarray*}


\noindent The second-order equation for the longitudinal coordinate is given
by

\begin{eqnarray}
\tau_{\text{2}}(s,\theta)&=&\sum_{1,2}\left[\sigma_{k}(s)\cos(k\theta)+\delta_{k}(s)\sin(k\theta)\right]
 + \tau_{2a}(s) 
 + \nonumber \\ &+&  
\left[\frac{2\left(p_{\text{x0a}}(s){}^{2}+p_{\text{y0a}}(s){}^{2}+1\right)+a_0^{2}(s)\left(m_{0}(s)^{2}+n_{0}(s)^{2}\right)}{2j_{0}(s)^{2}}
- 
\tau_{\text{1a}}'(s)-1 \right] \theta,\label{eq:tau2Solution}
\end{eqnarray}

\noindent where the coefficients of the oscillatory part are

\begin{eqnarray*}
\sigma_{1}(s)&=&\frac{2a_{0}(s)n_{0}(s)p_{\text{x0a}}(s)}{j_{0}^{2}(s)}, \quad
\sigma_{2}(s)=\frac{a_{0}^{2}(s)n_{0}(s)m_{0}(s)}{2j_{0}^{2}(s)}, \\
\delta_{1}(s)&=&\frac{2a_{0}(s)m_{0}(s)p_{\text{x0a}}(s)}{j_{0}^{2}(s)}, \quad
\delta_{2}(s)=\frac{a_{0}^{2}(s)\left[m_{0}^{2}(s)-n_{0}^{2}(s)\right]}{2j_{0}^{2}(s)},
\end{eqnarray*}


\noindent The longitudinal momentum is found to be 

\begin{eqnarray}
p_{\text{z1}}(s,\theta)&=&
\frac{p_{\text{x0}}(s,\theta) p_{\text{x1}}(s,\theta)+p_{\text{y0}}(s,\theta) p_{\text{y1}}(s,\theta)}{j_{0}(s)}   + \nonumber \\ 
&+& \frac{\gamma_{0}(s,\theta) 
\left[ 
\Pi(s)j_{0}(s) + 
a_0(s)\left(
m_{1}(s)\sin\theta
 + n_{1}(s)\cos\theta 
\right) 
\right]}{j_{0}(s)}
+ 
\theta j_{0}'(s)/j_{0}(s). \label{eq:pz1Solution}
\end{eqnarray}


\noindent The functions $x_{1a}(s)$, $y_{1a}(s)$, $p_{x1a}(s),$ $p_{y1a}(s)$,
$\tau_{2a}(s)$, and $\Pi(s)$ would have to be calculated from higher-order
equations, but the non-secular behaviour conditions necessary to completely
define the ground-state can be drawn from Eqs. (\ref{eq:x1Solution})-(\ref{eq:pz1Solution}).
The prerequisites obviously read

\begin{eqnarray}
p_{\text{x0a}}(s) &=& j_{0}(s)x_{0a}'(s),\qquad p_{\text{y0a}}(s)=j_{0}(s)y_{0a}'(s)   \label{eq:SecularityRemovalTransverseCoords} \\
p_{\text{x0a}}'(s) &=& -\frac{a_{0}^{2}(s)\left(m_{0}(s)m_{1}(s)+n_{0}(s)n_{1}(s)\right)}{2j_{0}(s)},  \label{eq:SecularityRemovalMomentaIntermeiateX} \\
p_{\text{y0a}}'(s) &=& -\frac{a_{0}^{2}(s)\left(m_{0}(s)m_{2}(s)+n_{0}(s)n_{2}(s)\right)}{2j_{0}(s)}, \label{eq:SecularityRemovalMomentaIntermeiateY} \\
j_{0}(s) &=& const,   \label{jConstpz1Solution} \\
\tau_{\text{1a}}'(s)& =&
\frac{ 
 a_{0}^2(s)\left(m_{0}^2(s)+n_{0}^2(s)\right)    
 + 
 2 \left(p_{\text{x0a}}^2(s)+p_{\text{y0a}}^2(s)+1 -j_{0}^2(s)\right)}{2 j_{0}^2(s)} 
 .    \label{eq:FirstOrderLongitudinalCoordEq}
\end{eqnarray}

Eqs. (\ref{eq:SecularityRemovalMomentaIntermeiateX}) and   
(\ref{eq:SecularityRemovalMomentaIntermeiateY}) can be cast in the
form

\begin{eqnarray}
p_{\text{x0a}}'(s)+\frac{a_{0}^{2}(s)}{2j_{0}}\partial_{x}W(x_{0a}(s),y_{0a}(s),\tau_{0a})&=&0,\label{eq:SecularityRemovalTransverseMomentumX} \\
p_{\text{y0a}}'(s)+\frac{a_{0}^{2}(s)}{2j_{0}}\partial_{y}W(x_{0a}(s),y_{0a}(s),\tau_{0a})&=&0,\label{eq:SecularityRemovalTransverseMomentumY} \\
W(x,y,\tau)=\frac{m^{2}(x,y,\tau)+n^{2}(x,y,\tau)}{2}&=&\frac{\left|u(x,y,\tau)\right|}{2}^{2},\label{eq:PonormdtivePotentialGeneralForm}
\end{eqnarray}

\noindent meaning that $W_{p}(x,y,\tau,s)=a_{0}^{2}(s)W(x,y,\tau)$ plays the role of the ponderomotive potential of the original system. 
Eqs. (\ref{eq:SecularityRemovalTransverseCoords})-(\ref{jConstpz1Solution}), (\ref{eq:SecularityRemovalTransverseMomentumX}), and (\ref{eq:SecularityRemovalTransverseMomentumY})
represent the ponderomotive dynamics problem for an electron in the relativistic case and a rigorous generalization of the classic result of \cite{GaponovMiller}.

\section{Averaged Dynamics in Cylindrical Coordinates}

Consider Eqs. (\ref{eq:SecularityRemovalTransverseCoords})-(\ref{jConstpz1Solution}) 
in cylindrical coordinates, the transition being introduced by $x_{0a}(s)=r(s)\cos(\varphi(s))$,
$y_{0a}(s)=r(s)\sin(\varphi(s))$, $p_{\text{x0a}}(s)=\rho(s)\cos(\psi(s))$,
$p_{\text{y0a}}(s)=\rho(s)\sin(\psi(s))$. Further denote $\delta\psi(s)=\varphi(s)-\psi(s)$
and

\[
W(x,y,\tau)=c\left(r^{2},\varphi,\tau_{0a}\right).
\]
 Eqs. (\ref{eq:SecularityRemovalTransverseCoords})-(\ref{jConstpz1Solution})
thereby translate into 

\begin{eqnarray}
r'(s)&=&\frac{\rho(s)\cos(\delta\psi(s))}{j_{0}},\label{eq:CircularSymmRadEq} \\
\rho'(s)&=&\frac{a_{0}^{2}(s)]}{4j_{0}r(s)} \left[\sin\left(\delta\psi(s)\right)\partial_{\varphi}c\left(r^{2}(s),\varphi(s),\tau_{0a}\right) 
 - 2\cos\left(\delta\psi(s)\right)r^{2}(s)\frac{\partial c\left(r^{2}(s),\varphi(s),\tau_{0a}\right)}{\partial r^{2}}\right], \nonumber \\
\varphi'(s)&=&-\frac{\sin(\delta\psi(s))\rho(s)}{j_{0}r(s)}, \nonumber \\ 
\psi'(s)&=&-\frac{a_{0}^{2}(s)}{4j_{0}r(s)\rho(s)} \left[\cos\left(\delta\psi(s)\right) 
\frac{\partial c\left(r^{2}(s),\varphi(s),\tau_{0a}\right)}{\partial \varphi} 
+ 
2\sin\left(\delta\psi(s)\right)r^{2}(s)\frac{\partial}{\partial r^{2}}c\left(r^{2}(s),\varphi(s),\tau_{0a}\right)\right]. \nonumber
\end{eqnarray}

The equations in the cylindrical coordinate frame further allow for a reduction  to
a problem with three unknowns instead of four if the field intensity
is axially symmetric. In this case, we denote $c\left(r^{2},\varphi,\tau_{0a}\right)=c_{0}\left(r^{2},\tau_{0a}\right)$
and arrive at a set of electron dynamics equations comprising Eq.
(\ref{eq:CircularSymmRadEq}) and

\begin{eqnarray}
\rho'(s)&=&-\frac{a_{0}^{2}(s)r(s)\cos\left(\delta\psi(s)\right)\frac{\partial}{\partial r^{2}}c_0\left(r^{2}(s),\tau_{0a}\right)}{2j_{0}},\label{eq:CircularSymmMomentumEq} \\
\delta\psi'(s)&=&\frac{\sin\delta\psi(s)\left[a_{0}^{2}(s)r^{2}(s)\frac{\partial c_0\left(r^{2}(s),\tau_{0a}\right)}{\partial r^{2}}
 -
2\rho^{2}(s)\right]}{2j_{0}r(s)\rho(s)}.\label{eq:CircularSymmPhaseDiffEq}
\end{eqnarray}
In particular, 
$c_{0}\left(r^2,\tau_{0a}\right)=\exp\left[-2r^{2}/\left(\tau_{0a}^{2}+1\right)\right]/\left(\tau_{0a}^{2}+1\right)$ for a Gaussian pulse.

In what follows, the model for the laser pulse longitudinal profile
is $a_{0}(s)=q\exp[-(s-d)^{2}/\sigma^{2}]$,
where $q$ is the pulse peak amplitude, $\sigma$ is the pulse duration,
and $d$ is the distance between the pulse peak and the initial coordinate
of the electron.

\section{Ponderomotive Symmetry Breaking \\ in an Intense Optical Field}

Considering the interaction of the optical field with a sufficiently sparse target consisting
of a large number of electrons, it is natural to
assume that the electrons sustain unidirectionally random motion prior
to the target's being overrun by the propagating electromagnetic
pulse. In this case, the initial values of momentum angle $\psi$ uniformly span
the whole range from 0 to 2$\pi$ for every initial value of the coordinate
angle $\varphi$. Therefore, the scattering of the above ensemble is axially symmetric, and Eqs. (\ref{eq:CircularSymmRadEq})-(\ref{eq:CircularSymmPhaseDiffEq})
should be treated for consecutive values of $\delta\psi$ similarly varying over the full angular difference range.
Two representative cases of the electron scattering by a laser pulse
are shown below. In the first one, depicted in Fig. 1, the laser radiation
disperses a target initially located at a large distance from the
pulse peak. The scattering occurs chiefly at the pulse front, which
explains the electron energy outputs appearing to be modest against
the ultrarelativistic laser intensity (note that this circumstance
was observed both in experiments and in prior simulations based on direct
solution of the electron dynamics equations \cite{MyLPB2015,LPB-2017}).
In the second case, illustrated in Fig. 2, the electrons are initially
placed within the domain occupied by the optical field as in the case
of the ionization self-injection \cite{HuStarace1,HuStarace2} (no
attempt is made in this study to simulate the ionization dynamics
consistent with the laser field). Under the arrangement, the electrons
exposed to the central part of the laser pulse are released with much
higher energies than in the case shown in Fig. 1. The initial conditions in all calculations are formulated
so that electrons take off with relatively small random momenta which
correspond to the adiabatic invariant $j_{0}$ exhibiting slight random
deviations from unity.

The paradigm exemplified by both cases is the symmetry breaking in
the dynamics of the electrons driven by the transversely localized
optical field. The phase difference $\delta\psi$ vanishes quickly
and the ponderomotively pressured electron gains momentum on the
average to be, in the long run, ejected from the interaction zone.
Calculations demonstrate that the release of electrons covered by
the focal spot, as evidenced by Figs. 1 and 2, occurs almost exactly
at the focal spot edge defined by $\left(r(s)/\sqrt{1+\tau^{2}}\right)=1$. The electron ejection angle inferred from Eqs. (\ref{eq:AdiabaticInvariant})
and (\ref{eq:GroundStateLongitudinalMomentum}) is expressed as $\theta_{\text{eject}}=\tan^{-1}\left[ 2j_{0}\rho/ \left(1+\rho^{2}-j_{0}^{2}\right)\right]$.

The scattering of an ensemble of electrons is further examined on the basis of the proposed theory. The energy-directionality diagram for the same optical field parameters as in the case of Fig. 1 is presented in Fig. 3. The numbers of electrons within specific ranges of angles relative to the field propagation direction are displayed, with the angles expressed in terms of their energy equivalents as prescribed by Eq. (\ref{eq:GroundStateLongitudinalMomentum}) and the ensuing expressions for the relativistic mass factor and energy. Sub-ponderomotive noise, i.e. the energies having magnitudes which border on the precision of the asymptotic method employed, are filtered from the plot. Fig. 4 demonstrates in detail two representative energy spectra resulting from the scattering of electrons into specific angular ranges. An energy-directionality diagram and two representative electron energy spectra for the field parameters of Fig. 2 are shown in Figs. 5 and 6 respectively. 

\begin{figure}[h] 
\includegraphics[width=0.99\textwidth]{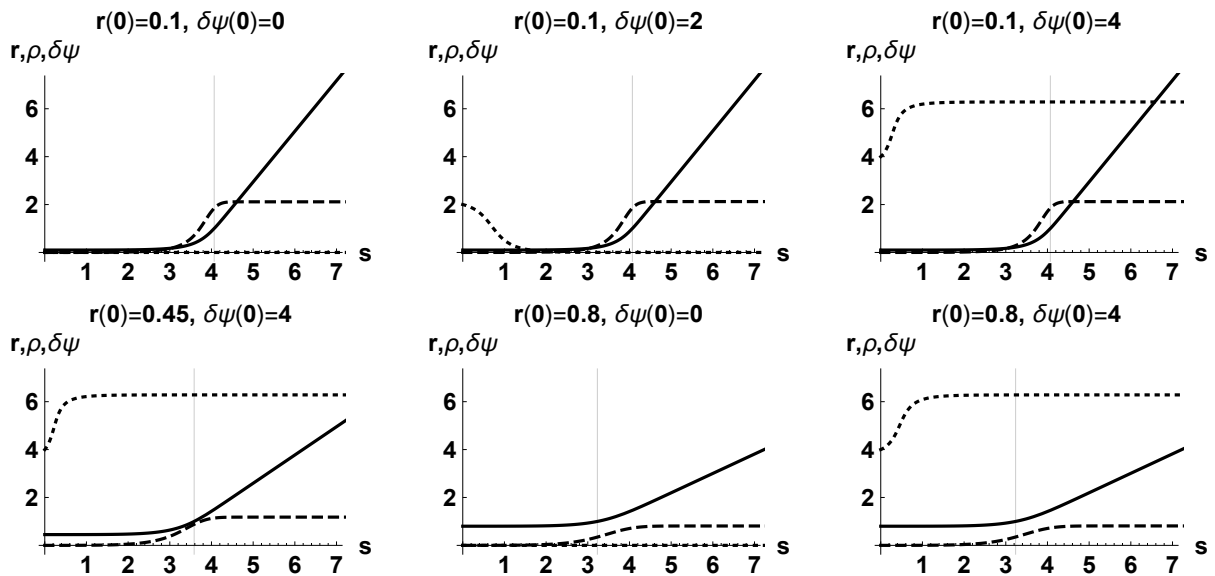}
\caption{Averaged radial coordinate $r(s)$ (continuous line),
radial component of momentum $\rho(s)$ (dashed line), and phase difference
between coordinate and momentum $\delta\psi(s)$ (dotted line)
for an electron driven by a relativistically intense focused optical
envelope. The optical field parameters are $q=$33, $\sigma=$4, $d=$10 (interaction with a distant target).
Computation results are shown for $\tau_{0a}=$0 (focal plane) and
a range of initial positions of an electron in this plane. The envelope
small parameter is $\epsilon=$0.1. The initial electron momenta are
generated as random values on the order of 0.1\% of the relativistic
threshold (the adiabatic invariant fluctuates from case to case accordingly).
The results are depicted in proper time and show clearly the steady
gain of energy on the average by the electron and the ultimate release
of the electron from the state of being captured by the field due
to the symmetry breaking in the focused electromagnetic envelope.
}
\end{figure}

\begin{figure}[h] 
\includegraphics[width=0.99\textwidth]{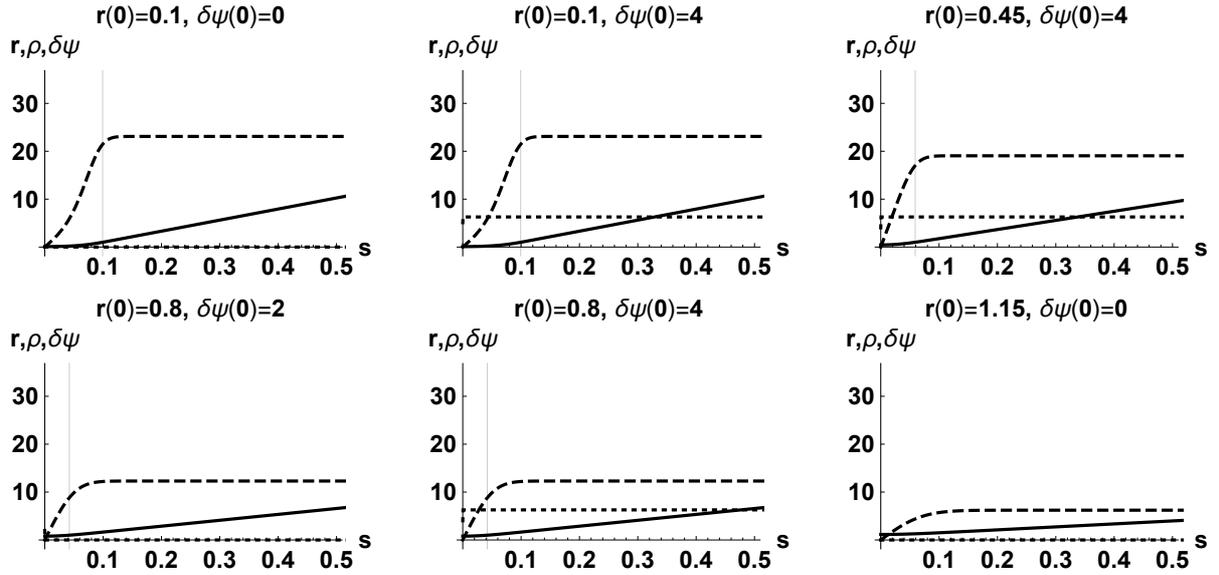}
\caption{Averaged radial coordinate $r(s)$ (continuous line),
radial component of momentum $\rho(s)$ (dashed line), and phase difference
between coordinate and momentum $\delta\psi(s)$ (dotted line)
for an electron driven by a relativistically intense focused optical
envelope. The optical field parameters
are $q=$33, $\sigma=$4, $d=$0 (self-injection). Computation results
are shown for $\tau_{0a}=$0 (focal plane) and a range of initial
positions of an electron in this plane. The envelope small parameter
is $\epsilon=$0.1. The initial electron momenta are generated as
random values on the order of 0.1\% of the relativistic threshold
(the adiabatic invariant fluctuates from case to case accordingly).
The results are depicted in proper time and show clearly the steady
gain of energy on the average by the electron and the ultimate release
of the electron from the state of being captured by the field due
to the symmetry breaking in the focused electromagnetic envelope.}
\end{figure}

\begin{SCfigure}  
\includegraphics[width=0.5\textwidth]{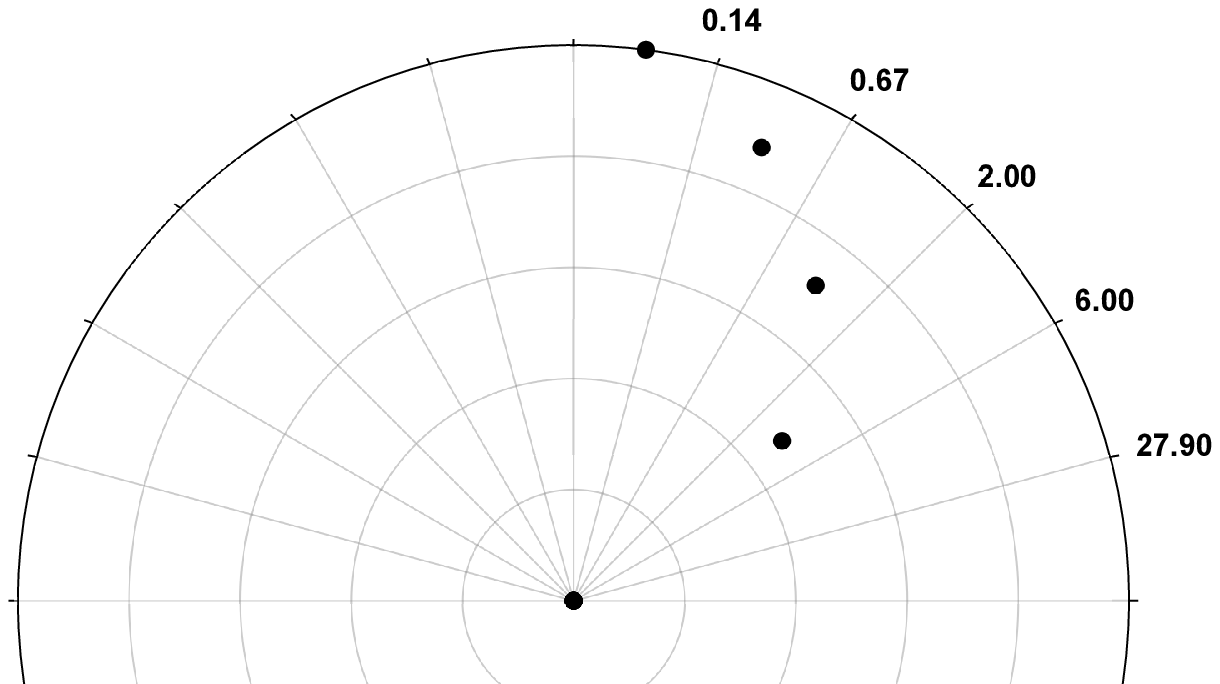}
\caption{Diagrams of electron scatter directionality relative
to the optical field propagation axis for the optical field parameters cited
in Fig. 1. The sub-ponderomotive noise is filtered out and the angles
are expressed in terms of the electron energy estimates based on the adiabatic invariant
with $j_{0}=1$. The actual electron
energies may spread slightly across the limits shown due to
individual fluctuations of the value of $j_{0}$ per individual particle,
while electrons with some of the energies marginally fitting into
the specified energy ranges may be absent.}
\end{SCfigure}

\begin{figure} 
\includegraphics[width=0.99\textwidth]{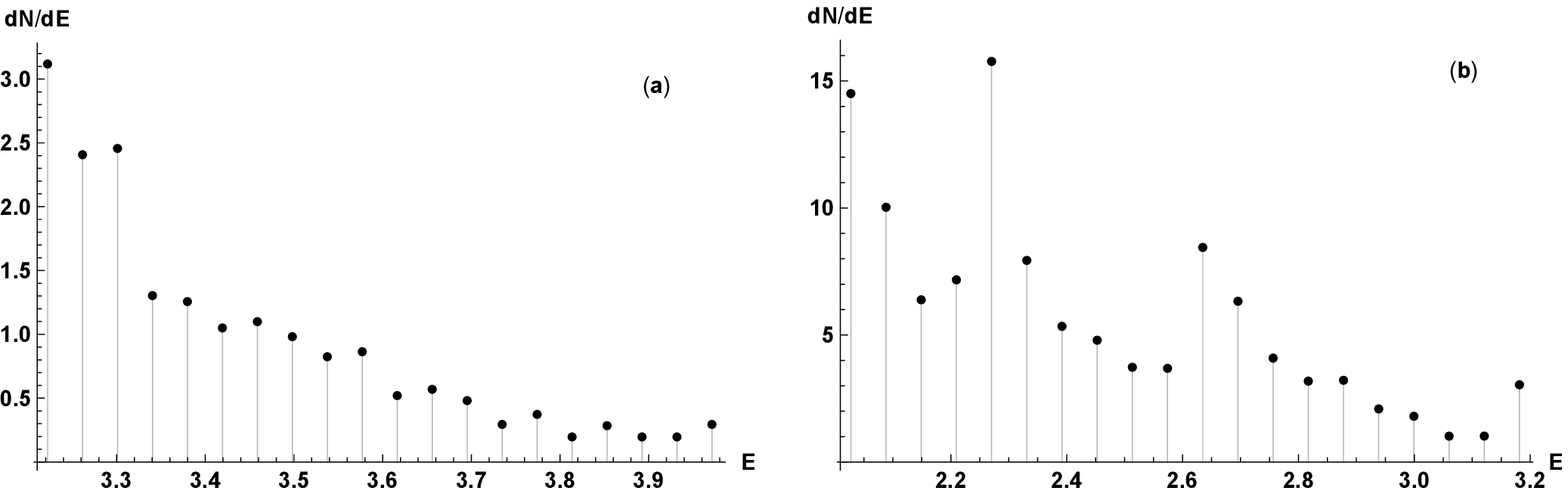}
\caption{Electron release energies for the optical field parameters cited
in Fig. 1. (a) Electron scatter energy spectrum for 0.59$\leq\theta\leq$0.67.
(b) Electron scatter energy spectrum for 0.67$\leq\theta\leq$0.79.
}
\end{figure}

\begin{SCfigure} 
\includegraphics[width=0.5\textwidth]{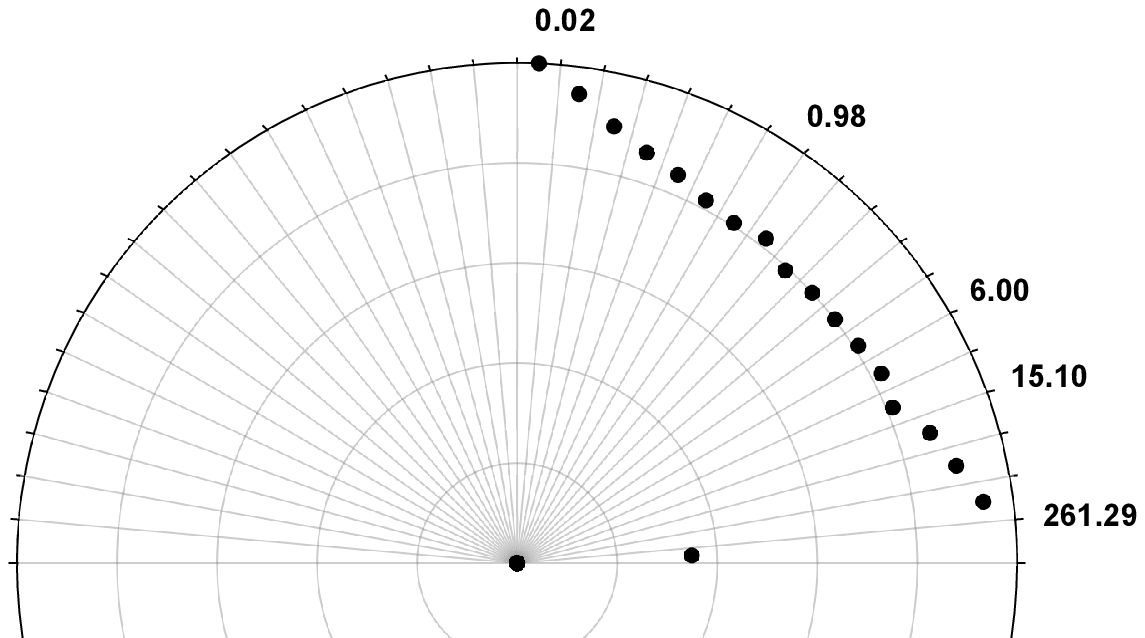}
\caption{Diagrams of electron scatter directionality relative
to the optical field propagation axis for the optical field parameters cited
in Fig. 2. The sub-ponderomotive noise is filtered out and the angles
are expressed in terms of the electron energy estimates based on the adiabatic invariant
with $j_{0}=1$. The actual electron
energies may spread slightly across the limits shown due to
individual fluctuations of the value of $j_{0}$ per individual particle,
while electrons with some of the energies marginally fitting into
the specified energy ranges may be absent.}
\end{SCfigure}

\begin{figure} 
\includegraphics[width=0.99\textwidth]{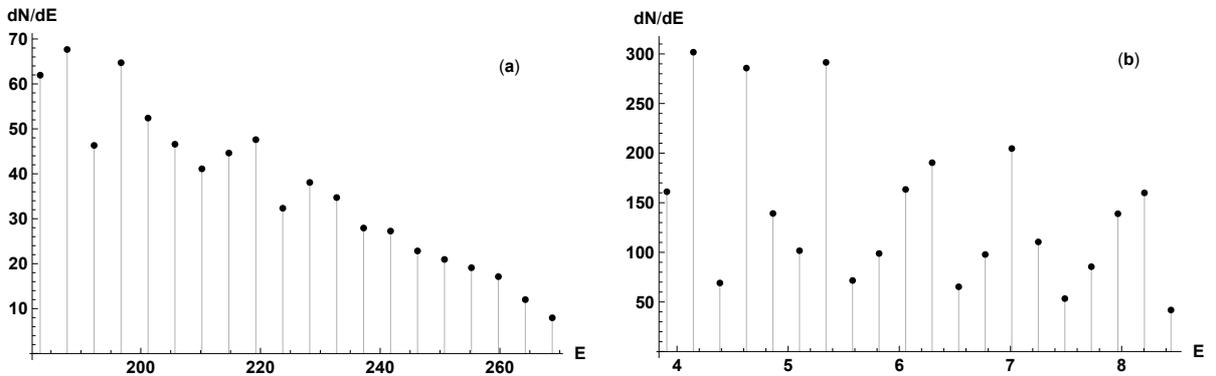}
\caption{Electron release energies for the optical field parameters cited
in Fig. 2. (a) Electron scatter energy spectrum for 0.08$\leq\theta\leq$0.1.
(b) Electron scatter energy spectrum for 0.45$\leq\theta\leq$0.63.
}
\end{figure}

\section{Conclusions}

A rigorous asymptotic approach is employed to describe the relativistic
ponderomotive effects in the dynamics of an electron driven by an
intense focused finite-duration electromagnetic envelope, the expansion
parameter being proportional to the ratio of the field wavelength
to the focal spot size. The resulting theory of ponderomotive dynamics
is further applied to model the scattering of a sparse ensemble of
electrons by a relativistically intense laser pulse. A refined set
of electromagnetic field equations is suggested to consistently account
for the envelope shape, the longitudinal component of the propagating
pulse, and the field corrections due to finite-duration effects. Approximate
solutions to the relativistic equations of the electron dynamics are
generated with the help of the Krylov-Bogolyubov technique which leads
to a set of equations de facto averaged over field oscillations and
written in terms of the proper time of the electron, as well as to
the establishment of an adiabatic invariant linking the parameters
of the post-interaction motion picture for an electron in proper time
to its ejection angle and release energy in the original reference
frame. For pulses with axially symmetric intensity profiles, the relativistic
ponderomotive equations are found to be reducible to a form which
immediately affords averaging out the random angular dependencies
in initial conditions for the electron ensemble. Electron scatter
directionality diagrams and energy spectra within selected angles
are calculated from the resulting compact model.

\end{document}